# Dynamic strain in gold nanoparticle supported graphene induced by focused laser irradiation

András Pálinkás[1,2], Péter Kun[1,2], Antal A. Koós[1,2], Zoltán Osváth[1,2,]*

[1]*Institute of Technical Physics and Materials Science, MFA, Centre for Energy Research, Hungarian Academy of Sciences, 1525 Budapest, P.O. Box 49, Hungary*

[2]*Korea-Hungary Joint Laboratory for Nanosciences (KHJLN), P.O. Box 49, 1525 Budapest, Hungary*

## Abstract

Graphene on noble-metal nanostructures constitutes an attractive nanocomposite with possible applications in sensors or energy conversion. In this work we study the properties of hybrid graphene/gold nanoparticle structures by Raman spectroscopy and Scanning Probe Methods. The nanoparticles (NPs) were prepared by local annealing of gold thin films using focused laser beam. The method resulted in a patterned surface, with NPs formed at arbitrarily chosen microscale areas. Graphene grown by chemical vapour deposition was transferred onto the prepared, closely spaced gold NPs. While we found that successive higher intensity (6 mW) laser irradiation increased gradually the doping and the defect concentration in $SiO_2$ supported graphene, the same irradiation procedure did not induce such irreversible effects in the graphene supported by gold NPs. Moreover, the laser irradiation induced dynamic hydrostatic strain in the graphene on Au NPs, which turned out to be completely reversible. These results can have implications in the development of graphene/plasmonic nanoparticle based high temperature sensors operating in dynamic regimes.

*Corresponding author. E-mail address: zoltan.osvath@energia.mta.hu

## Introduction

Graphene has been in the focus of unceasing interest for both fundamental and applied research due to its unique material properties[1]. However, assigning functionality and integrating graphene into nanodevices has been challenged by the necessity of reliable and accessible preparation and characterization methods. Scanning probe microscopy and Raman spectroscopy are frequently used characterization methods for graphene and other two-dimensional (2D) materials. Moreover, Raman-spectroscopy has been settled as a standard tool[2,3] to measure the quality of graphene, such as the number of layers[4], the doping[5], the strain[6,7,8,9,10] in graphene, etc. As a standard, it has been considered a non-invasive method, which is true only for low power measurements. The laser power can alter the samples in various ways[11,12], especially if used in ambient conditions, for example by laser induced thermal oxidation[13,14,15,16] or increase in chemical doping[17]. On the other hand, when determining a safe power density, it should be taken into account that recording Raman spectra with too low laser power leads to bad signal-to-noise ratio. Oxidation might occur even at very low laser power densities, which are often regarded non-invasive. According to Kraus et al.[11], this occurs on a time scale of few hours for cleaved graphene, while only a few tens of seconds for graphene grown by chemical vapour deposition (CVD)[12]. The laser induced changes during Raman spectroscopy investigations of graphene has been investigated comprehensively mostly with $SiO_2$ as substrate. Although this is a common substrate used in the field, it is of immediate importance to reveal the possible effects that can occur on other prevalent supporting materials[18], now including different nanoparticles (NPs).

Graphene on noble-metal nanostructures has been demonstrated to be an attractive nanocomposite with many possible applications[19,20,21]. The surface-enhanced Raman spectroscopy (SERS) on noble-metal NPs is widely used to measure very low amounts of molecules, in some cases even single molecules can be detected[22]. Graphene-enhanced Raman scattering, which is considered to be based on a chemical enhancement mechanism[23], also has significant potential in microanalysis[24,25]. Combining the two processes, i.e. the SERS of the noble-metal NPs and the chemical enhancement of graphene could even improve the possibilities, either by depositing plasmonic structures onto graphene[26,27,28,29], or by transferring graphene onto the metal NPs[30,31,32]. In this latter case the graphene shields the metal NPs from chemical interactions[33], while its chemisorption activity increases due to the corrugated nature of these substrates[34,35]. The Raman signal of graphene itself can also be amplified by annealing, as the suspended graphene regions fill better the space between the NPs where the plasmonic enhancement of the electric field is the highest. As graphene follows the shape of the NPs, strain will emerge, and the improved adherence will modify also the electrostatic doping. This process is traceable with Raman spectroscopy and atomic force microscopy (AFM) measurements[31]. In this work we study the effects of optically induced heating of graphene-gold hybrid nanostructures. We produce gold NPs locally by focused laser irradiation of gold thin films. We show that the shift of Raman peaks of graphene transferred onto the NPs can be attributed mainly to strain, which is switched on and off by the applied laser.

Experimental

Graphene-gold hybrid nanostructures were prepared as follows: gold grains of 99.99999 % purity were applied as source material for evaporation, at a background pressure of $3\times10^{-8}$ mbar. A thin gold film of 5 nm was deposited onto a $SiO_2$(285 nm)/Si substrate by an electron-beam evaporation system (AJA) at a rate of 0.2 nm s$^{-1}$. Gold nanoparticles were formed on areas of 5×5 µm$^2$ by local laser heating of the gold layer using a confocal Raman microscope (WITec) and a focused, 6 mW laser-power @633 nm. The laser beam was focused into a wavelength-wide spot and scanned over the selected 5×5 µm$^2$ regions with 20×20 points and a dwelling time of 3 seconds in each point. Commercially available CVD graphene from Graphenea Inc. was transferred onto the substrate (gold layer and gold NPs) using thermal release tape (TRT) method as described elsewhere[31]. For comparison, we prepared a sample where graphene was transferred directly to a standard, SiO$_2$(285 nm)/Si substrate.

The samples were investigated by confocal Raman microscopy (WITec) using excitation lasers of 488 and 633 nm. Low laser powers (0.6 mW) were used to characterize the same areas both before and after local heating. Raman maps were recorded in order to study the spatial distribution of spectral peaks. In this technique the excitation laser is scanned in a defined geometry and a complete Raman spectrum is recorded in every stepping point. AFM measurements were performed on a MultiMode 8 (Bruker) operating in tapping mode with SuperSharpSilicon™ probes (NANO*SENSORS*, tip radius ~2 nm). The sample with graphene on the 5 nm gold film was studied by scanning tunnelling microscopy (STM) and tunnelling spectroscopy (STS), using a DI Nanoscope E operating under ambient conditions. The STM measurements were performed in constant current mode with mechanically-cut Pt/Ir (90/10%) tips.

## Results and discussion

The thin gold film (5 nm) evaporated on the SiO$_2$(285 nm)/Si substrate was locally transformed into dome-like gold NPs by systematically annealing[36] with a focused laser beam, as described above. The transformation can be easily observed optically, since the reflectance of the irradiated layer changed, as shown in Fig. 1a. The AFM image in Fig. 1b shows both the as-evaporated gold layer (left side) and one of the irradiated parts (right side) with the borderline in the middle. Dome-like gold NPs formed with diameter of 32 nm and height of 11 nm in average. The distribution of the maximum height of more than 2000 particles can be seen in Fig. 1c (red). Here, the height distribution of the as-evaporated layer with a sharp peak at 5 nm is also shown (black).

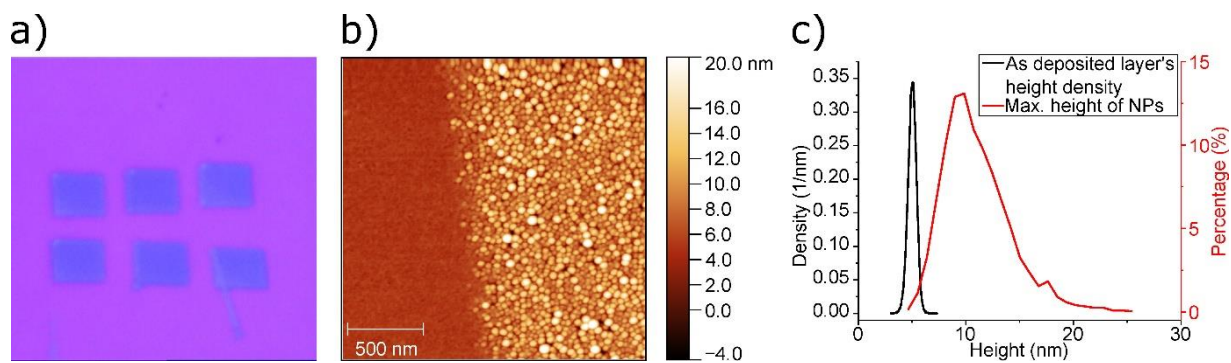

*Figure 1.* a) Optical image after 6mW laser annealing. The irradiated areas (blue) can be well distinguished from the non-irradiated areas (purple). b) AFM image showing the border of as-evaporated (left) and annealed region (right). c) Height distribution of the non-irradiated area of b) (black line) and the maximum height distribution of 2460 gold NPs formed on the irradiated area in b) (red line).

We transferred CVD-grown graphene onto the prepared gold nanoparticles. Fig. 2a shows an AFM-image of a typical laser irradiated, 5×5 µm² region and its surroundings, after the transfer process. Here graphene covers most of the area. Several wrinkles can be observed and also a discontinuity in graphene (darker contrast) as a result of breaking during the transfer. Multiple Raman-spectroscopy measurements were performed on this region and on the surroundings with low (0.6 mW) and high (6 mW) laser power, as described later. Note that the higher laser power caused the formation of the nanoparticles from the evaporated gold layer. We selected two regions to demonstrate the effect of the high laser power on the sample: the first is near a corner-like graphene edge (Fig. 2b-c, red dashed line) on a region with NPs already formed, while the second region includes the graphene-covered border between the nanoparticles and the continuous film (Fig. 2d-e, purple dashed line). These two regions were marked in Fig. 2a with blue and green squares, respectively. Fig. 2b and 2c show AFM images of the same corner-like graphene edge, right after the transfer process and after multiple high power laser annealing, respectively. The initially formed nanoparticles remain unchanged after subsequent laser annealing (see for example the NPs marked with black crosses in Fig. b-c, and the configuration of the NPs surrounding them). We can observe in Fig. 2b that the covering graphene, is rippled in various elongated shapes. In turn, in Fig. 2c the discrete nanoparticles are well outlined beneath the graphene, because graphene follows better the shape of NPs after laser annealing (see also Fig. S1, Supplementary Information). To further illustrate the effect of laser annealing, let us consider Fig. 2d and 2e, where we marked the edges of the graphene (red dashed lines), the borderline of the area with NPs (purple dashed line) and a large graphene wrinkle (black arrow). Subsequent high power laser irradiation of the whole area shown in Fig. 2d resulted in the formation of gold NPs beneath the graphene as well (Fig. 2e, upper part of the image, above the purple dashed line), and eventually no difference can be found between the two regions initially separated by the borderline (Fig. 2e). Note that the large graphene wrinkle is ironed out during the process, and graphene mimics the corrugation of the underlying NPs.

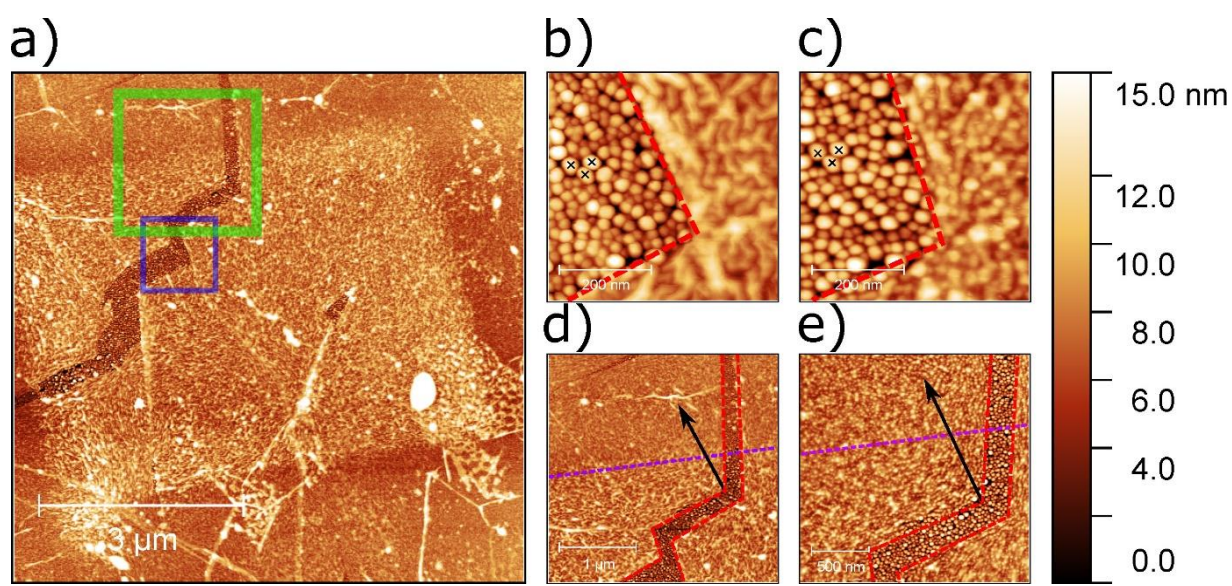

*Figure 2.* *The structure of the graphene-gold NPs nanocomposite revealed by AFM. a) Graphene covers most of the area. Wrinkles, holes, and cracks can be noticed. The areas marked with blue and green squares are detailed in b)-c) and d)-e), respectively. b)-c) AFM images showing both graphene covered and bare gold NPs, b) before, and c) after subsequent laser annealing. The corner-like graphene edge is marked with red dashed line, while three individual bare gold NPs are marked with black symbols. d)-e) AFM images where we marked the edges of the graphene (red dashed lines), the border of initially irradiated area (purple dashed line) and a large graphene wrinkle (black arrow). Subsequent laser annealing (e) produces NPs under graphene, and irons out the wrinkle.*

In the following we demonstrate the effect of laser irradiation by analysing the Raman peaks of both graphene on standard $SiO_2$ substrate and on gold NPs. For this, we used correlation diagrams of the spectral positions of graphene G and 2D peaks, suggested by Lee et al.[37] to separate the mechanical strain from charge doping in graphene. It has to be noted, that in the original approach the initial strain configuration (uniaxial or biaxial) and the doping type (p- or n-type) of graphene have to be known in order to obtain a good separation. Nevertheless, in the recent work of N. S. Mueller et al.[3], the authors introduce a new method where one can separate the effect of the hydrostatic and shear strain from doping without any initial assumption on the strain configuration. We plotted the slopes of the pure hydrostatic strain and the pure p-type doping with gradients of $(\Delta\omega_{2D}/\Delta\omega_G)_\epsilon^{hydrostatic}$=2.21[3] and $(\Delta\omega_{2D}/\Delta\omega_G)_n^{hole}$=0.55[38], respectively. We used the Grüneisen parameter of $\gamma_G$=1.8 given by Zabel et al.[39], which is in good agreement with theoretical calculations[6]. We want to underline that the hydrostatic strain is twice as large as the corresponding biaxial strain. For hole doping, the dependence of the G peak position is taken as $\Delta Pos(G)/\Delta E_F \approx 39$ cm$^{-1}$/eV [5]. The equilibrium values at room temperature are marked with light green square symbols in the Figures: $(\omega_G; \omega_{2D})$ = (1582 cm$^{-1}$; 2691 cm$^{-1}$)[40] for excitation with a 488 nm wavelength laser and $(\omega_G; \omega_{2D})$ = (1581 cm$^{-1}$; 2635 cm$^{-1}$) for the 633 nm wavelength excitation. First, we performed successive Raman maps with 488 nm laser excitation on a SiO$_2$/Si supported graphene in a 5×5 µm$^2$ area with 20×20 pixels. The excitation laser power was measured right before the experiments. Low (0.6 mW) and high (6 mW) power measurements were performed alternately. The higher laser power was used explicitly to anneal locally the samples, while with low powers we characterized the effect of the applied heating. It is important to note that the laser annealing always takes place in areas of only about 0.5 × 0.5 µm$^2$ (spots) at the same time, were the laser is focused. As the laser beam is scanned above the sample, previously irradiated spots quickly cool down back to room temperature.

In Fig. 3a we plotted the 2D-G peak positions, where we used different colours for the subsequent measurements on the same area. The average (G,2D) position from the first, low-power measurement (black dots) exhibit a small hole doping of -18±13 meV and a compressive hydrostatic strain of -0,06±0,01%, which are reasonable values for CVD graphene on SiO$_2$ substrate. The second (red dots) and the forth (light blue dots) measurements were performed with high laser power, while the third (blue dots) and fifth (green dots) were again low-power measurements. The corresponding average strain and doping values are -0,07±0,01%; -140±34 meV and -0,07±0,01%; -238±19 meV for the third and the fifth measurement, respectively. A significant increase in the p-type doping is clearly seen from the correlation plot, the shift of the averages follow the pure p-doping slope of 0.55 very well. The average spectra from the first and fifth measurements are shown in Fig. 3b, where one can notice – apart from the peak shifts – the lowered intensity ratio of the G-2D peaks, which is also a fingerprint of the increased doping. The defect related D peak slightly increased, which shows that the high power laser irradiation not only heats the sample but also introduces defects. These defects, as well as the structural deformations induced by laser annealing, capture more easily airborne contaminants[16], giving a plausible explanation for the observed overall increase in doping.

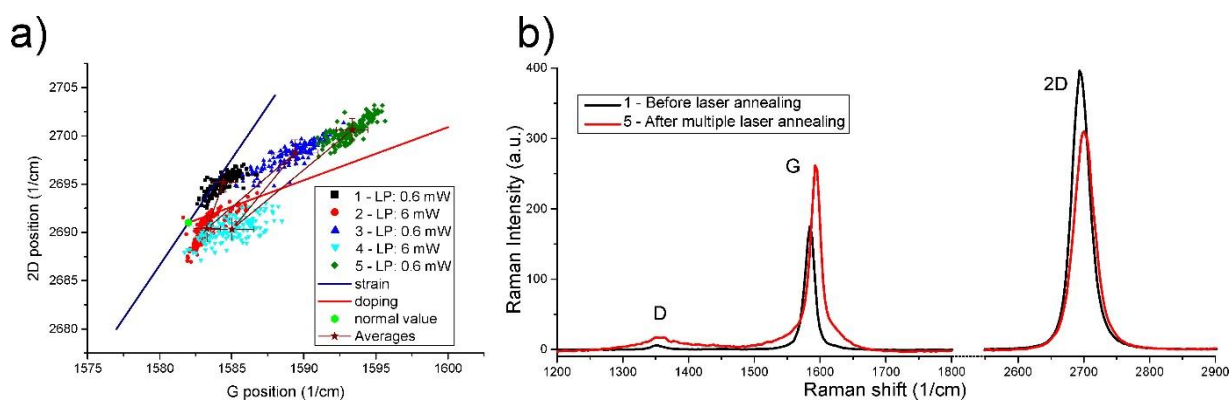

*Figure 3.* Correlation plot of the 2D-G Raman peaks' position of Si/SiO$_2$ supported graphene measured with 488 nm laser excitation. a) Subsequent Raman maps on the same area recorded with laser power of 0.6 mW (1$^{th}$, 3$^{rd}$, 5$^{th}$) or 6 mW (2$^{nd}$, 4$^{th}$). b) Average of the Raman spectra from the first and the last map shows shifted peaks with changed relative intensity.

Now we turn the attention to the graphene supported by gold nanoparticles. Similarly, we performed Raman measurements with low laser power (0.6 mW) and high laser power (6 mW) alternately, and scanning successively an area of 5×5 μm$^2$ of the graphene/Au NPs nanocomposite (see Fig. S2-c, Supplementary Information). In Fig. 4 we plotted the 2D-G correlation plots (a, c) and the average spectra (b, d) for laser excitations of 633 and 488 nm, respectively. For clarity, we normalised the spectra to the Si peak (at 520 1/cm) for the spectra measured with 488 nm excitation, and to the graphene 2D peak for the spectra measured with 633 nm excitation. The first noticeable difference is the presence of the photoluminescent (PL) background of the metal nanoparticles. First we discuss this phenomenon, because it provides a standalone evidence on the temperature and the volume change, and will be helpful in the interpretation of graphene Raman spectra. When the excitation was performed with 2.54 eV photons (488 nm), we measured a complete fluorescence spectrum, while in the case of 1.95 eV photons (633 nm) we observed only the lower, decreasing part of the spectrum. This is due to the fact that the red laser only captures the low energy part of the induced interband transitions[41]. The temperature increase due to the laser irradiation is clearly seen both on the redshift and on the lowered intensity of the fluorescent background. The corresponding maximum of 1861 1/cm (536 nm) shifts to 2380 1/cm (552 nm) when the laser power is switched from low to high, as shown in Fig. 4d. Such redshifts and lowered intensity in PL and surface plasmon resonance frequencies is attributed to thermal dilatation, which also means a decrease of electron density[41,42,43].

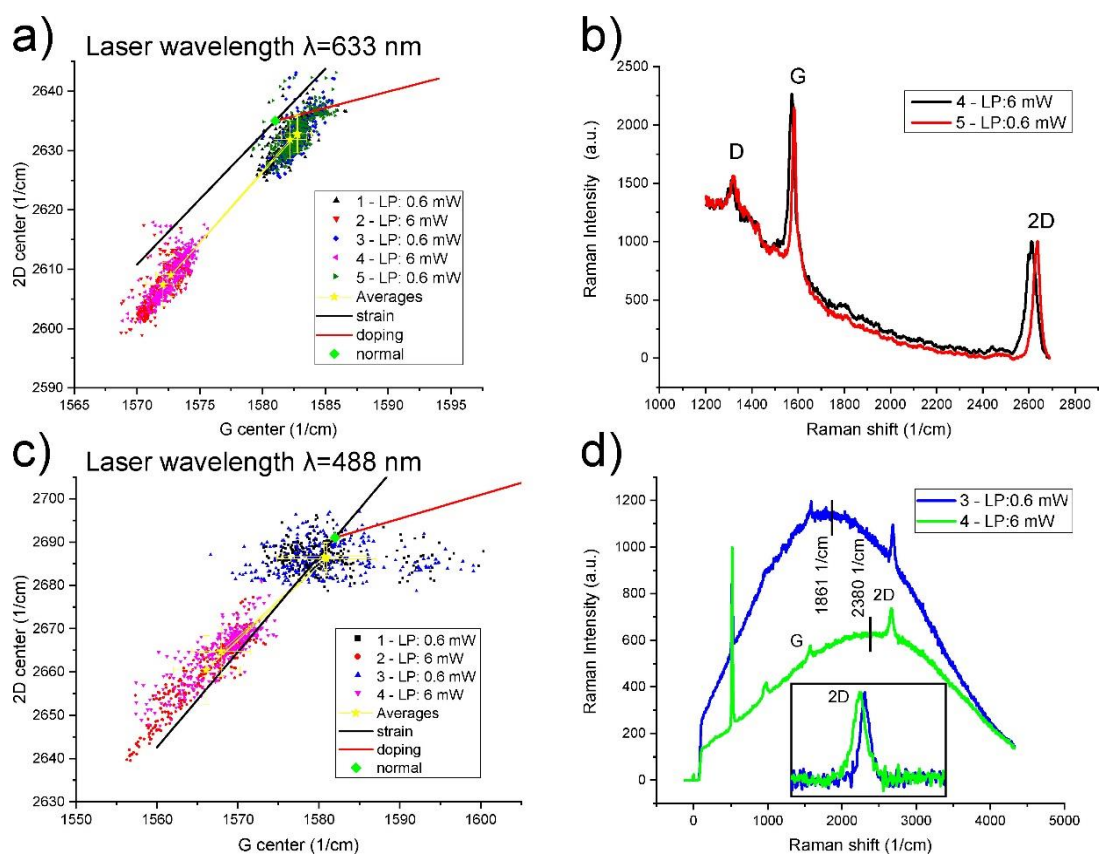

*Figure 4.* 2D-G correlation plots of Raman peak positions of the Au NPs supported graphene measured with 633 nm (a) and 488 nm (c) laser excitation. Average spectra are shown in (b) and (d), respectively. Successive measurements are numbered and marked with coloured symbols. Low- (0.6 mW) and high (6 mW) laser power (LP) measurements were performed alternately. Background subtracted 2D peaks are shown in the inset of d).

On the average spectra one can notice the redshift of the graphene Raman peaks for high laser power. More insights can be revealed from the analysis of the correlation plots. In the case of the 633 nm excitation (Fig. 4a) the average of the first low power measurement was at ($\omega_G$; $\omega_{2D}$)=(1582.2; 2631.9), which corresponds to a small tensile strain $\varepsilon^{hidr}_1$ = 0.08% and p-doping of $E_{F1}$ = - 89 meV. The following high power measurements (no. 2 and 4) show a significant shift of both G and 2D peaks, with the averages ($\omega_G$; $\omega_{2D}$)=(1572.1; 2607.4) and ($\omega_G$; $\omega_{2D}$)=(1572.7; 2609) for measurement no. 2 and 4, respectively. Note that the effect is reproducible, the peak shifts are very similar in the two cases. Furthermore, the low power measurements are also very reproducible, the no. 3 and 5 have averages close to the first one. For the blue laser we found the same behaviour. On average, neither the strain nor the doping was changed significantly, as shown in Fig. 4c. We should note, however, that here the standard deviation of the G peak position was very large, probably due to the high PL background. In order to analyse the G peak shifts $\Delta\omega_G$ in Figs. 4a and 4c, one should consider the contributions from thermal expansion of the lattice ($\Delta\omega_G^E$), anharmonic effect ($\Delta\omega_G^A$), and strain induced by the thermal expansion coefficient (TEC) mismatch between the substrate and graphene ($\Delta\omega_G^S$)[44,45]:

$$\Delta\omega_G(T) = \Delta\omega_G^E(T) + \Delta\omega_G^A(T) + \Delta\omega_G^S(T), \quad (1)$$

where $T$ is the temperature of the sample induced by high power laser irradiation. The contribution from strain can be written as

$$\Delta\omega_G^S(T) = \beta \int_{300K}^{T} (\alpha_{Au\ NP}(T) - \alpha_{graphene}(T))dT, \quad (2)$$

where β is the biaxial strain coefficient of the G band[9], $\alpha_{Au\ NP}(T)$ and $\alpha_{graphene}(T)$ are the temperature dependent TECs of the gold NPs and graphene, respectively. A prominent feature observed on the correlation plots (Figs. 4a and 4c) is the nearly parallel shift of the Raman-peaks with respect to the pure hydrostatic strain slope. We fitted the average values for each laser and found the slopes to be 2.37±0.02 and 1.72±0.07 for the 633 nm and for the 488 nm laser, respectively, which are close to 2.2 (the "strain" slope)[38]. This clearly shows that the main origin of the shift is due to hydrostatic strain from the underlying Au NPs, expressed by Eq. (2). As we discussed above, the thermal expansion of the gold nanoparticles can be observed from the shift of the PL background. Taking this into account together with the negative thermal expansion coefficient of graphene[44,46] it is straightforward that a tensile strain will emerge in graphene. This result is in agreement with recent findings showing the dominant role of thermal strain in the temperature dependence of graphene Raman peaks[45,47]. Nevertheless, the temperature dependence of phonon properties remains a challenging topic, as it is highly affected by the supporting materials and their TECs [[45,47]]. In Fig. 4, if we assume p-type doping during the whole low-high laser power cycle, the strain is dynamically increased from 0.08±0.01% to 0.48±0.06%, and from 0.07±0.02% to 0.46±0.11% during red and blue laser irradiation, respectively. Each spectral point from the measurements no. 2 and 4 (for both lasers) are aligned along the pure hydrostatic strain slope, which shows that the emerging strain varies in a very wide scale as a result of the relatively broad distribution of nanoparticle heights and diameters. Another interesting aspect of the measurements is that no observable doping or damaging occurred in graphene, even after several power cycles, unlike on the $SiO_2$ substrate (Fig. 3), which is rather surprising. The 2D peak width analysis confirms this finding, there is no significant change in the full width at half maximum (FWHM) distributions corresponding to the successive low power measurements (see Supplementary Information, Figs. S3 and S4). Most of the reports from the literature[5,11–16] which investigate the resistance of graphene to radiation defects used standard Si/$SiO_2$ substrate and found similar behaviour in terms of increased damaging and doping. It was pointed out that the presence of water can seriously affect the photo-oxidation of graphene[15]. It is known that a thin water layer can be trapped between the silicon-dioxide and the transferred graphene, and in most of the cases a water film could form on the top of graphene as well[48,49,50], although there is a debate on the wetting properties of graphene[51,52,53,54]. We suppose that the graphene on Au is more hydrophobic than graphene on $SiO_2$[51], and suggest that the absence of water layer (graphene on Au NPs) helps the graphene to resist the high power laser irradiation. Further investigations on the role of the supporting material regarding the resistance of graphene to photoradiation would be useful.

Finally, the structure and local electronic properties of graphene on the continuous Au film were investigated by STM and STS. An STM image recorded on a fully graphene covered region of the 5 nm thin gold layer is shown in Fig. 5a. Interestingly, the corrugation of graphene is quite similar to the initial corrugation of graphene transferred onto Au NPs (Fig. 2b). This is related to the morphology of the as deposited 5 nm gold layer (Fig. S2, Supplementary Information), on which graphene is partially suspended. Extended graphene ripples can be observed, similar to the ripples observed on non-annealed graphene by AFM (Fig. 2b). STS measurements were performed in several arbitrary points to investigate the local doping of graphene. Fig. 5b shows four average spectra corresponding to four different spots on the sample. The spectra show slightly varying doping, with the Dirac point observed between 66 meV and 94 meV. This is in very good agreement with the result obtained from the Raman spectroscopy of Au NPs supported graphene (89 meV, black dashed line in Fig. 5b).

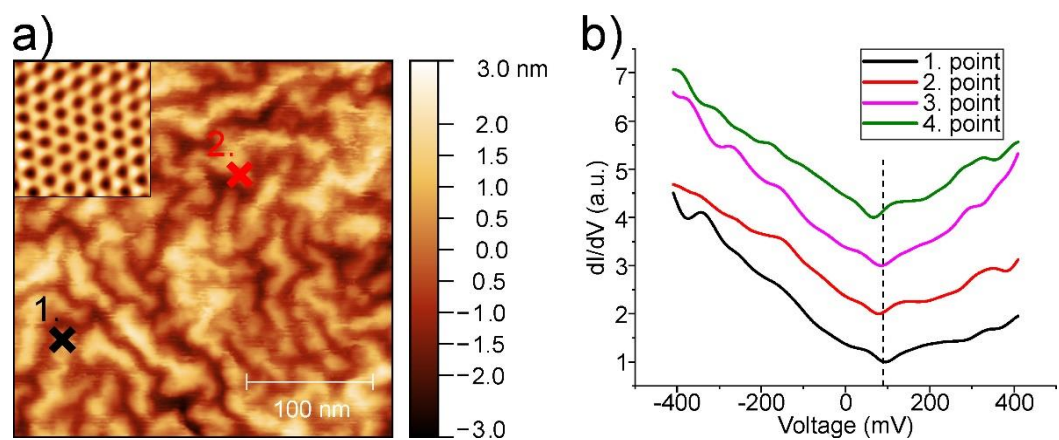

*Figure 5. a) STM topography image of graphene covered thin gold layer, measured in constant current mode (U = -50 mV, I = 0.5 nA). The atomic lattice of graphene is shown in the inset. b) STS spectra measured in different locations of the sample, two of which are marked in a). The position of the Dirac point as obtained from Raman spectroscopy is marked with black dashed line.*

## Conclusions

Gold nanoparticles were produced locally by focused laser irradiation of thin Au films evaporated on SiO$_2$/Si substrate. The properties of graphene transferred onto these NPs were studied by AFM, STM/STS, and confocal Raman spectroscopy. We demonstrated that dynamic strain could be induced in the Au NP supported graphene by high power (6 mW) laser irradiation, an effect which was completely reversible upon switching off the laser. We also showed that – while similar laser irradiation induced increased doping and damage in SiO$_2$/Si supported graphene – no change in doping or defect concentration was observed on Au NP supported graphene, even after several irradiation cycles. Our findings can gain importance in local-heat assisted applications like plasmonic sensors, spasers, or photothermal therapy with nanoparticles or graphene/nanoparticle hybrids. Since the high temperature associated with high power laser irradiation increases the strain in graphene, which in turn enhances its sensing properties, these results can have implications in the development of graphene/plasmonic nanoparticle based high temperature sensors as well, operating either in steady-state or dynamic regimes.

## Conflicts of interest

There are no conflicts of interest to declare.

## Acknowledgements

The Authors acknowledge financial support from the National Research, Development and Innovation Office (NKFIH) in Hungary, through the OTKA Grant K 119532, and from the Korea-Hungary Joint Laboratory for Nanosciences. Z.O. expresses his thanks to Á. Juhász for the production of the table of contents graphic.

## Notes and references

[1] E.P. Randviir, D.A.C. Brownson, C.E. Banks, A decade of graphene research: production, applications and outlook, Mater. Today 17 (2014) 426–432. doi:10.1016/j.mattod.2014.06.001

[2] A.C. Ferrari, D.M. Basko, Raman spectroscopy as a versatile tool for studying the properties of graphene, Nat. Nanotechnol. 8 (2013) 235–246. doi:10.1038/nnano.2013.46.

[3] N.S. Mueller, S. Heeg, M.P. Alvarez, P. Kusch, S. Wasserroth, N. Clark, F. Schedin, J. Parthenios, K. Papagelis, C. Galiotis, M. Kalbáč, A. Vijayaraghavan, U. Huebner, R. Gorbachev, O. Frank, S. Reich, Evaluating arbitrary strain configurations and doping in graphene with Raman spectroscopy, 2D Mater. 5 (2018) 15016. doi:10.1088/2053-1583/aa90b3.

[4] L.M. Malard, M.A. Pimenta, G. Dresselhaus, M.S. Dresselhaus, Raman spectroscopy in graphene, Phys. Rep. 473 (2009) 51–87. doi:10.1016/j.physrep.2009.02.003.

[5] M. Bruna, A.K. Ott, M. Ijäs, D. Yoon, U. Sassi, A.C. Ferrari, Doping dependence of the Raman spectrum of defected graphene, ACS Nano. 8 (2014) 7432–7441. doi:10.1021/nn502676g.

[6] F. Ding, H. Ji, Y. Chen, A. Herklotz, K. Dörr, Y. Mei, A. Rastelli, O.G. Schmidt, Stretchable graphene: A close look at fundamental parameters through biaxial straining, Nano Lett. 10 (2010) 3453–3458. doi:10.1021/nl101533x.

[7] T.M.G. Mohiuddin, A. Lombardo, R.R. Nair, A. Bonetti, G. Savini, R. Jalil, N. Bonini, D.M. Basko, C. Galiotis, N. Marzari, K.S. Novoselov, A.K. Geim, A.C. Ferrari, Uniaxial strain in graphene by Raman spectroscopy: G peak splitting, Grüneisen parameters, and sample orientation, Phys. Rev. B - Condens. Matter Mater. Phys. 79 (2009). doi:10.1103/PhysRevB.79.205433.

[8] C. Neumann, S. Reichardt, P. Venezuela, M. Drögeler, L. Banszerus, M. Schmitz, K. Watanabe, T. Taniguchi, F. Mauri, B. Beschoten, S. V. Rotkin, C. Stampfer, Raman spectroscopy as probe of nanometre-scale strain variations in graphene, Nat. Commun. 6 (2015) 1–7. doi:10.1038/ncomms9429.

[9] D. Yoon, Y.W. Son, H. Cheong, Strain-dependent splitting of the double-resonance raman scattering band in graphene, Phys. Rev. Lett. 106 (2011). doi:10.1103/PhysRevLett.106.155502.

[10] M. Mohr, J. Maultzsch, C. Thomsen, Splitting of the Raman 2D band of graphene subjected to strain, Phys. Rev. B - Condens. Matter Mater. Phys. 82 (2010). doi:10.1103/PhysRevB.82.201409.

[11] B. Krauss, T. Lohmann, D.H. Chae, M. Haluska, K. Von Klitzing, J.H. Smet, Laser-induced disassembly of a graphene single crystal into a nanocrystalline network, Phys. Rev. B - Condens. Matter Mater. Phys. 79 (2009) 1–9. doi:10.1103/PhysRevB.79.165428.

[12] G. Amato, G. Milano, U. Vignolo, E. Vittone, Kinetics of defect formation in chemically vapor deposited (CVD) graphene during laser irradiation: The case of Raman investigation, Nano Res. 8 (2015) 3972–3981. doi:10.1007/s12274-015-0900-1.

[13] A.E. Islam, S.S. Kim, R. Rao, Y. Ngo, J. Jiang, P. Nikolaev, R. Naik, R. Pachter, J. Boeckl, B. Maruyama, Photo-thermal oxidation of single layer graphene, RSC Adv. 6 (2016) 42545–42553. doi:10.1039/C6RA05399H.

[14] F. Herziger, R. Mirzayev, E. Poliani, J. Maultzsch, In-situ Raman study of laser-induced graphene oxidation, Phys. Status Solidi Basic Res. 252 (2015) 2451–2455. doi:10.1002/pssb.201552411.

[15] N. Mitoma, R. Nouchi, K. Tanigaki, Photo-oxidation of graphene in the presence of water, J. Phys. Chem. C. 117 (2013) 1453–1456. doi:10.1021/jp305823u.

[16] S. Ryu, L. Liu, S. Berciaud, Y.J. Yu, H. Liu, P. Kim, G.W. Flynn, L.E. Brus, Atmospheric oxygen binding and hole doping in deformed graphene on a SiO2 substrate, Nano Lett. 10 (2010) 4944–4951. doi:10.1021/nl1029607.

[17] L.M. Malard, R.L. Moreira, D.C. Elias, F. Plentz, E.S. Alves, M. Pimenta, Thermal enhancement of chemical doping in graphene: A Raman spectroscopy study, J. Phys. Condens. Matter. 22 (2010). doi:10.1088/0953-8984/22/33/334202.

[18] N. Papasimakis, S. Mailis, C.C. Huang, F. Al-Saab, D.W. Hewak, Z. Luo, Z.X. Shen, Strain engineering in graphene by laser irradiation, Appl. Phys. Lett. 106 (2015). doi:10.1063/1.4907776.

[19] I. Khalil, N.M. Julkapli, W.A. Yehye, W.J. Basirun, S.K. Bhargava, Graphene-gold nanoparticles hybrid-synthesis, functionalization, and application in a electrochemical and surface-enhanced raman scattering biosensor, 2016. doi:10.3390/ma9060406.

[20] Z. Fang, Y. Wang, Z. Liu, A. Schlather, P.M. Ajayan, F.H.L. Koppens, P. Nordlander, N.J. Halas, Plasmon-induced doping of graphene, ACS Nano. 6 (2012) 10222–10228. doi:10.1021/nn304028b.

[21] Y. Wu, W. Jiang, Y. Ren, W. Cai, W.H. Lee, H. Li, R.D. Piner, C.W. Pope, Y. Hao, H. Ji, J. Kang, R.S. Ruoff, Tuning the doping type and level of graphene with different gold configurations, Small. 8 (2012) 3129–3136. doi:10.1002/smll.201200520.

[22] S. Nie, S.R. Emory, Probing single molecules and single nanoparticles by surface-enhanced Raman scattering, Science (80-.). 275 (1997) 1102–1106. doi:10.1126/science.275.5303.1102.

[23] H. Xu, L. Xie, H. Zhang, J. Zhang, Effect of graphene Fermi level on the Raman scattering intensity of molecules on graphene, in: ACS Nano, 2011: pp. 5338–5344. doi:10.1021/nn103237x.

[24] X. Ling, L. Xie, Y. Fang, H. Xu, H. Zhang, J. Kong, M.S. Dresselhaus, J. Zhang, Z. Liu, Can graphene be used as a substrate for Raman enhancement?, Nano Lett. 10 (2010) 553–561. doi:10.1021/nl903414x.

[25] S. Huh, J. Park, Y.S. Kim, K.S. Kim, B.H. Hong, J.M. Nam, UV/ozone-oxidized large-scale graphene platform with large chemical enhancement in surface-enhanced Raman scattering, ACS Nano. 5 (2011) 9799–9806. doi:10.1021/nn204156n.

[26] F. Schedin, E. Lidorikis, A. Lombardo, V.G. Kravets, A.K. Geim, A.N. Grigorenko, K.S. Novoselov, A.C. Ferrari, Surface-Enhanced Raman Spectroscopy of Graphene, ACS Nano. 4 (2010) 5617–5626. http://dx.doi.org/10.1021/nn1010842.

[27] J. Lee, S. Shim, B. Kim, H.S. Shin, Surface-Enhanced Raman Scattering of Single- and Few-Layer Graphene by the Deposition of Gold Nanoparticles, Chem. - A Eur. J. 17 (2011) 2381–2387. doi:10.1002/chem.201002027.

[28] R. Nicolas, G. Lévêque, P.M. Adam, T. Maurer, Graphene Doping Induced Tunability of Nanoparticles Plasmonic Resonances, Plasmonics. (2017) 1–7. doi:10.1007/s11468-017-0623-0.

[29] W. Xu, X. Ling, J. Xiao, M.S. Dresselhaus, J. Kong, H. Xu, Z. Liu, J. Zhang, Surface enhanced Raman spectroscopy on a flat graphene surface, Proc. Natl. Acad. Sci. 109 (2012) 9281–9286. doi:10.1073/pnas.1205478109.

[30] W. Xu, J. Xiao, Y. Chen, Y. Chen, X. Ling, J. Zhang, Graphene-veiled gold substrate for surface-enhanced raman spectroscopy, Adv. Mater. 25 (2013) 928–933. doi:10.1002/adma.201204355.

[31] Z. Osváth, A. Deák, K. Kertész, G. Molnár, G. Vértesy, D. Zámbó, C. Hwang, L.P. Biró, The structure and properties of graphene on gold nanoparticles, Nanoscale. 7 (2015) 5503–5509. doi:10.1039/C5NR00268K.

[32] S. Heeg, R. Fernandez-Garcia, A. Oikonomou, F. Schedin, R. Narula, S.A. Maier, A. Vijayaraghavan, S. Reich, Polarized plasmonic enhancement by Au nanostructures probed through raman scattering of suspended graphene, Nano Lett. 13 (2013) 301–308. doi:10.1021/nl3041542.

[33] V. Berry, Impermeability of graphene and its applications, Carbon N. Y. 62 (2013) 1–10. doi:10.1016/j.carbon.2013.05.052.

[34] D.W. Boukhvalov, M.I. Katsnelson, Enhancement of chemical activity in corrugated graphene, J. Phys. Chem. C. 113 (2009) 14176–14178. doi:10.1021/jp905702e.

[35] C. Si, Z. Sun, F. Liu, Strain engineering of graphene: a review, Nanoscale. 8 (2016) 3207–3217. doi:10.1039/C5NR07755A.

[36] K. Grochowska, G. Śliwiński, A. Iwulska, M. Sawczak, N. Nedyalkov, P. Atanasov, G. Obara, M. Obara, Engineering Au Nanoparticle Arrays on SiO2 Glass by Pulsed UV Laser Irradiation, Plasmonics. 8 (2013) 105–113. doi:10.1007/s11468-012-9428-3.

[37] J.E. Lee, G. Ahn, J. Shim, Y.S. Lee, S. Ryu, Optical separation of mechanical strain from charge doping in graphene., Nat. Commun. 3 (2012) 1024. doi:10.1038/ncomms2022.

[38] G. Froehlicher, S. Berciaud, Raman spectroscopy of electrochemically gated graphene transistors: Geometrical capacitance, electron-phonon, electron-electron, and electron-defect scattering, Phys. Rev. B 91 (2015). doi:10.1103/PhysRevB.91.205413.

[39] J. Zabel, R.R. Nair, A. Ott, T. Georgiou, A.K. Geim, K.S. Novoselov, C. Casiraghi, Raman spectroscopy of graphene and bilayer under biaxial strain: Bubbles and balloons, Nano Lett. 12 (2012) 617–621. doi:10.1021/nl203359n.

[40] I. Calizo, A.A. Balandin, W. Bao, F. Miao, C.N. Lau, Temperature Dependence of the Raman Spectra of Graphene and Graphene Multilayers, Nano Lett. 7 (2007) 2645. doi:10.1021/nl071033g.

[41] V. Amendola, R. Pilot, M. Frasconi, O.M. Maragò, A.M. Iatì, Surface plasmon resonance in gold nanoparticles: a review, J. Phys. Condens. Matter. 29 (2017) 203002. doi:10.1088/1361-648X/aa60f3.

[42] O.A. Yeshchenko, I.S. Bondarchuk, V.S. Gurin, I.M. Dmitruk, A. V Kotko, Temperature dependence of the surface plasmon resonance in gold nanoparticles, Surf. Sci. 608 (2013) 275–281. doi:10.1016/j.susc.2012.10.019.

[43] O.A. Yeshchenko, I.S. Bondarchuk, V. V Kozachenko, Sensing the temperature influence on plasmonic field of metal nanoparticles by photoluminescence of fullerene C60 in layered C60/Au system, J. Appl. Phys. 117 (2015) 153102. doi:10.1063/1.4918554.

[44] D. Yoon, Y.W. Son, H. Cheong, Negative thermal expansion coefficient of graphene measured by raman spectroscopy, Nano Lett. 11 (2011) 3227–3231. doi:10.1021/nl201488g.

[45] W. Wang, Q. Peng, Y. Dai, Z. Qian, S. Liu, Temperature dependence of Raman spectra of graphene on copper foil substrate, J. Mater. Sci. Mater. Electron. 27 (2016) 3888–3893. doi:10.1007/s10854-015-4238-y.

[46] S. Mann, R. Kumar, V.K. Jindal, Negative thermal expansion of pure and doped graphene, RSC Adv. 7 (2017) 22378–22387. doi:10.1039/C7RA01591G.

[47] J. Judek, A.P. Gertych, M. Krajewski, K. Czerniak, A. Łapińska, J. Sobieski, M. Zdrojek, Statistical analysis of the temperature dependence of the phonon properties in supported CVD graphene, Carbon N. Y. 124 (2017) 1–8. doi:10.1016/j.carbon.2017.08.029.

[48] F. Schedin, A.K. Geim, S.V. Morozov, E.W. Hill, P. Blake, M.I. Katsnelson, K.S. Novoselov, Detection of Individual Gas Molecules Adsorbed on Graphene., Nat. Mater. 6 (2007) 652–5. doi:10.1038/nmat1967.

[49] T. Georgiou, L. Britnell, P. Blake, R. V. Gorbachev, A. Gholinia, A.K. Geim, C. Casiraghi, K.S. Novoselov, Graphene bubbles with controllable curvature, Appl. Phys. Lett. 99 (2011) 2011–2014. doi:10.1063/1.3631632.

[50] A. Ott, I.A. Verzhbitskiy, J. Clough, A. Eckmann, T. Georgiou, C. Casiraghi, Tunable D peak in gated graphene, Nano Res. 7 (2014) 338–344. doi:10.1007/s12274-013-0399-2.

[51] J. Rafiee, X. Mi, H. Gullapalli, A. V. Thomas, F. Yavari, Y. Shi, P.M. Ajayan, N.A. Koratkar, Wetting transparency of graphene, Nat. Mater. 11 (2012) 217–222. doi:10.1038/nmat3228.

[52] R. Raj, S.C. Maroo, E.N. Wang, Wettability of graphene, Nano Lett. 13 (2013) 1509–1515. doi:10.1021/nl304647t.

[53] Z. Li, Y. Wang, A. Kozbial, G. Shenoy, F. Zhou, R. McGinley, P. Ireland, B. Morganstein, A. Kunkel, S.P. Surwade, L. Li, H. Liu, Effect of airborne contaminants on the wettability of supported graphene and graphite, Nat. Mater. 12 (2013) 925–931. doi:10.1038/nmat3709.

[54] M. Munz, C.E. Giusca, R.L. Myers-Ward, D.K. Gaskill, O. Kazakova, Thickness-Dependent Hydrophobicity of Epitaxial Graphene, ACS Nano. 9 (2015) 8401–8411. doi:10.1021/acsnano.5b03220.